\documentclass[preprint]{emulateapj}

\begin{document}

\title{{\em Hubble Space Telescope} observations of the afterglow, supernova and host galaxy associated with the extremely bright GRB 130427A}

\shorttitle{The supernova in GRB 130427A}

\author{A.~J. Levan\altaffilmark{1}, 
N.~R. Tanvir\altaffilmark{2}, A.~S. Fruchter\altaffilmark{3}, J. Hjorth\altaffilmark{4}, E. Pian\altaffilmark{5,6,7}, P. Mazzali\altaffilmark{8},   R.~A. Hounsell\altaffilmark{3}
D.~A. Perley\altaffilmark{9,10}, Z. Cano\altaffilmark{11}, J. Graham\altaffilmark{3},  S.~B. Cenko\altaffilmark{12}, J.~P.~U. Fynbo\altaffilmark{4}, C. Kouveliotou\altaffilmark{13}, 
A. Pe'er\altaffilmark{14}, K. Misra\altaffilmark{15}, K. Wiersema\altaffilmark{2}}

\email{a.j.levan@warwick.ac.uk}

\altaffiltext{1}{Department of Physics, University of Warwick,
  Coventry, CV4 7AL, UK } 

\altaffiltext{2}{Department of Physics and Astronomy, University of Leicester, University Road,
Leicester, LE1 7RH, UK } 
  
\altaffiltext{3}{Space Telescope Science Institute, 3700 San Martin Drive, Baltimore, MD 21218, USA} 
\altaffiltext{4}{Dark Cosmology Centre, Niels Bohr Institute, University of Copenhagen, Juliane Maries Vej 30, DK-2100 Copenhagen, Denmark} 
\altaffiltext{5}{INAF, Trieste Astronomical Observatory, via G.B. Tiepolo 11, 34143 Trieste, Italy} 
\altaffiltext{6}{Scuola Normale Superiore, Piazza dei Cavalieri 7, 56126 Pisa, Italy} 
\altaffiltext{7}{European Southern Observatory, Karl-Schwarzschild-Strasse 2, 85748 Garching bei M\"{u}nchen, Germany}   
\altaffiltext{8}{Astrophysics Research Institute, Liverpool John Moores University, IC2 Liverpool Science Park 146 Brownlow Hill, Liverpool L3 5RF, United Kingdom.}
\altaffiltext{9}{Department of Astronomy, California Institute of Technology, MC 249-17, 1200 East California Blvd., Pasadena, CA 91125, USA} 
\altaffiltext{10}{Hubble Fellow}
\altaffiltext{11}{Centre for Astrophysics and Cosmology, Science Institute, University of Iceland, Dunhagi 5, 107 Reykjavik, Iceland}
\altaffiltext{12}{Astrophysics Science Division, NASA Goddard Space Flight Center, Mail Code 661, Greenbelt, MD, 20771, USA}
\altaffiltext{13}{Science and Technology Office, ZP12, NASA/Marshall Space Flight Center, Huntsville, AL 35812, USA}
\altaffiltext{14}{Department of Physics, University College Cork, Cork, Ireland}
\altaffiltext{15}{Aryabhatta Research Institute of Observational Sciences, Manora Peak, Nainital - 263 002, India}

\begin{abstract}
We present {\em Hubble Space Telescope} ({\it HST})
observations of the exceptionally bright and luminous {\em Swift} gamma-ray burst, GRB 130427A. 
At $z=0.34$ this burst affords an excellent opportunity to study the supernova and host galaxy associated with an 
intrinsically extremely luminous burst  ($E_{iso} >10^{54}$ erg):  more luminous than any previous GRB with a spectroscopically
associated supernova. We
use the combination of the image quality, UV capability  and and invariant PSF of {\em HST} to provide the best possible separation of
the afterglow, host and supernova
contributions to the observed light $\sim$17 rest-frame days after the burst  utilising a host subtraction spectrum
obtained 1 year later. 
 ACS grism observations show that the associated supernova, SN~2013cq, has an overall spectral shape 
and luminosity similar to SN~1998bw
(with a photospheric velocity, $v_{ph} \sim 15,000$ km s$^{-1}$). 
The positions of the bluer features are better matched by the higher velocity SN~2010bh  ($v_{ph} \sim 30,000$ km s$^{-1}$), but
this SN is significantly fainter, and fails to reproduce the overall spectral shape, perhaps indicative of velocity structure in the ejecta. 
We find that the burst
originated $\sim$4~kpc from the nucleus of a moderately star forming (1 M$_{\odot}$ yr$^{-1}$), possibly 
interacting disc galaxy. The absolute magnitude, physical size and morphology of this galaxy, 
as well as the location of the GRB within it are also strikingly
similar to those of GRB980425/SN~1998bw.
The similarity of supernovae and environment from both the most luminous and least
luminous GRBs suggests broadly similar 
progenitor stars 
can create GRBs across six orders of magnitude in isotropic energy. 
\end{abstract}

\section{Introduction}

The connection between long duration gamma-ray bursts (LGRBs) and hydrogen poor type Ic supernovae has become well established based on the detection of spectroscopic signatures of these supernovae accompanying a handful of relatively local GRBs
\citep[e.g.,][]{hjorth03,stanek03,soderberg04,pian06,bufano2012}. The GRB-SNe sample increases when combined 
with a larger set of events which exhibit photometric signatures in their lightcurves, consistent with SNe Ic  \citep[see e.g.,][]{cano13}.
Although, these light curve ``humps"
are not uniquely diagnostic of the supernova type, and are
open to alternative interpretations, the
emerging scenario is that at the majority of long GRBs are associated with type Ic supernovae \citep[e.g.][]{cano13}.

However, the picture painted by observations of such GRB-SNe pairs has remained unsatisfactory in some respects.
On average, these local events differ substantially
from the majority of the GRB population in terms 
of energy release, with isotropic energy releases ($E_{iso}$) a factor of $10^2-10^4$ lower than
the bulk population \citep[e.g.][]{kaneko07}. Of the bursts with the strongest evidence for SNe, 
only GRB~030329, with $E_{iso} \sim 10^{52}$ erg appears close to a being a
classical cosmological long-GRB. Several local GRB/SNe pairs exhibit $\gamma$-ray emission of
extremely long duration \citep{campana06,starling11}, while
other very long events at larger redshift show little evidence for SNe \citep{levan13}. Indeed, these local, low
luminosity bursts have been suggested to arise from a very different physical mechanism than the classical 
bursts, such as relativistic shock break-out from the supernova itself \citep[e.g.][]{nakar12}. Such emission is difficult to
locate in more luminous GRBs due to a combination of distance and glare from the burst itself, although
evidence for possible shock break-out components has been found in some GRBs \citep{starling12,sparre12}. However,
the several order of magnitude difference in energy release between the local, low-luminosity and
cosmological, high luminosity GRBs could also be indicative
of rather different physical mechanisms at play. 
Given this, the nature of the connection between the most energetic GRBs and their supernovae
remains in urgent need of further study.

Here we report observations of the brightest (highest fluence) GRB detected in the past $\sim$20 years, GRB 130427A. 
The isotropic energy release of $E_{iso} \sim 10^{54}$ erg, places
it in the most luminous 5\% of GRBs observed to date by {\em Swift}, and a factor of 100 brighter than GRB 030329 \citep{hjorth03} which
was the most luminous GRB with a well studied supernova.
At a redshift
of $z=0.340$ \citep{levan13b,xu13,perley13} the burst is close enough that any supernova is open to spectroscopic study, 
and indeed
 the presence of a supernova, SN~2013cq, has been established \citep{deugartepostigo13,xu13,melandri14}. 
 Here we use 
 the resolution of  the {\em Hubble Space Telescope} to resolve and dramatically reduce the galaxy contribution, and
 its UV capability to track the afterglow, hence enabling a view of the supernova as free as possible from the host, afterglow and atmospheric hinderance. 
  
\section{Observations}
GRB~130427A was discovered by {\em Swift} at 07:47:57 UT on 27 April 2013 \citep{maselli13}. 
It was also detected as an exceptionally bright GRB by {\em Konus-WIND} and {\em Fermi} with GBM \citep{fermi_gbm} and
LAT \citep{zhu}, and its prompt fluence of $S\approx 3 \times 10^{-3}$ ergs cm$^{-2}$  in the 10-1000 keV 
band \citep{fermi_gbm} makes it the most fluent GRB observed by {\em Swift}, Fermi or BATSE. It showed a bright X-ray and optical afterglow, peaking at R=7.4 before the {\em Swift} GRB trigger \citep{raptor}. Early spectroscopy of the afterglow yielded a redshift of $z=0.340$ \citep{levan13b}, which
was confirmed from later, more detailed spectroscopic observations \citep{xu13}. A full description of the afterglow is given in \cite{perley13,maselli14}. 
Deep photometric and spectroscopic observations over the first 10 days post burst 
revealed a re-brightening, consistent with the presence of a type Ic supernova, SN~2013cq \citep{deugartepostigo13,xu13}.

\begin{figure*}[ht]
    \centering
    \includegraphics[width=10cm,angle=0]{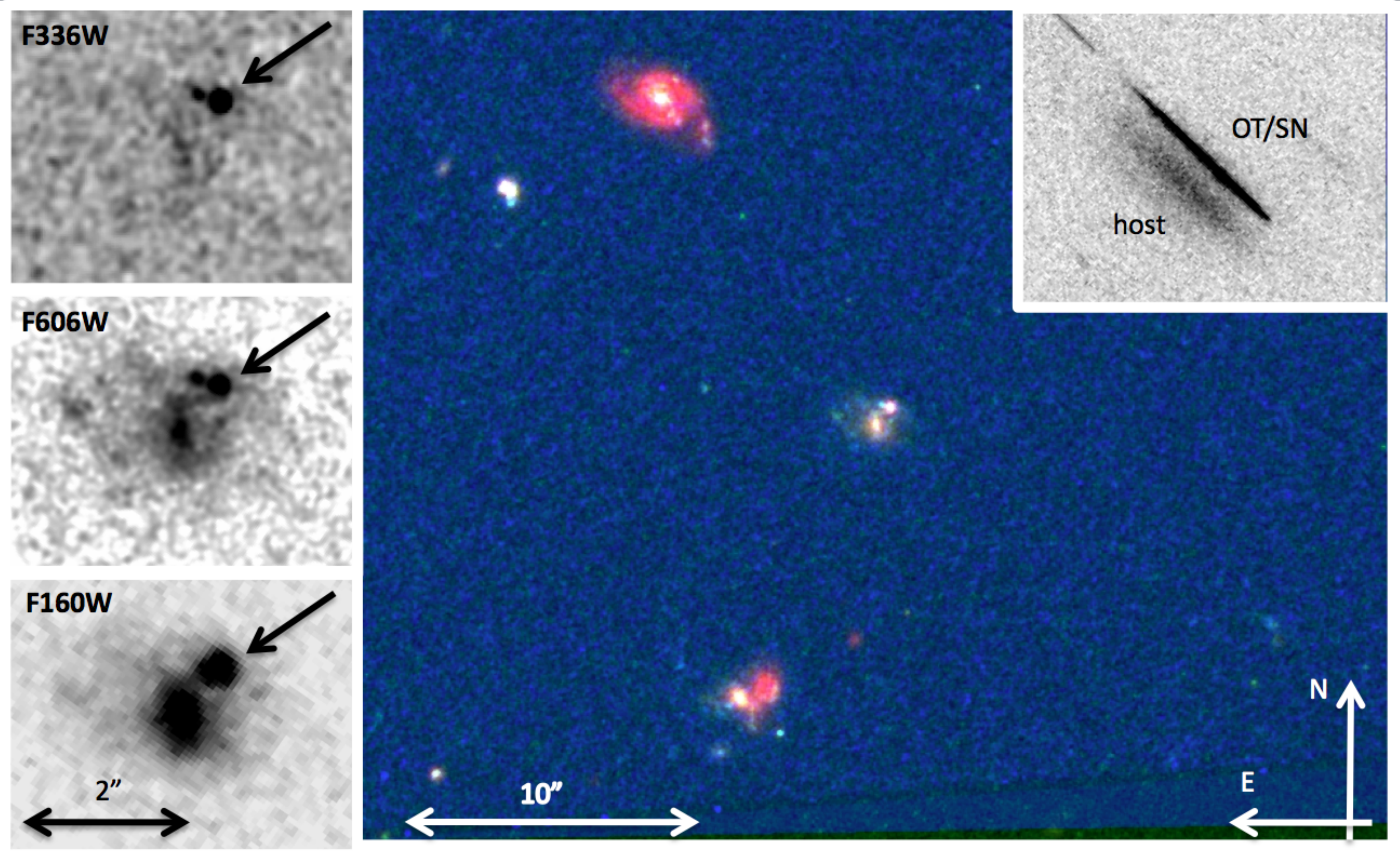}
\caption{Our {\em HST} observations of GRB~130427A. The left hand panel shows our UV-optical and IR imaging (the UV data taken on 20 May 2013, and the optical/IR on 10 July 2013). The afterglow, indicated by an arrow can be clearly seen offset 0.83\arcsec from the centre
of its host galaxy. In the UV the host is weakly detected, with a strong star forming region seen to the east of the GRB location. The F606W
image shows a disk galaxy with hints of a bar structure, along with some sign of distortion possibly due to ongoing interaction. The large
central panel shows a colour image from the three-band {\em HST} imaging. The host is at the centre, while other galaxies,
possibly part of a structure at the same redshift are visible. The top right hand panel shows our grism spectroscopy
of the host galaxy with the counterpart dispersed away from the host galaxy. }
\end{figure*}

\begin{deluxetable*}{lllllll} 
\tablecolumns{8} 
\tablewidth{0pc}
\tablecaption{Log of {\em HST} observations of GRB~130427A} 
\tablehead{ 
\colhead{Date-obs} & \colhead{MJD-obs} & \colhead{$\Delta T$ (d)} & \colhead{Filter} & \colhead{exp. (s)} & \colhead{AB-magnitude} }
\startdata
20-May-2013 & 56432.10521 & 22.78 & F336W & 1266  & 23.28 $\pm$ 0.02 \\ 
20-May-2013 & 56432.12503 & 22.80 & F160W & 1048 &  21.60 $\pm$ 0.01 \\ 
20-May-2013 &56432.37207 & 23.05 & F606W & 180  &  21.76 $\pm$ 0.01 \\ 
20-May-2013 & 56432.37603 &23.05 & G800L & 1880  & - \\
12-Apr-2014 & 56759.14866 & 349.82 & F336W & 2508 &26.17 $\pm$ 0.09 \\
18-Apr-2014 & 	56765.18459 &355.86  & F606W & 180 & 25.67 $\pm$ 0.11 \\ 
18-Apr-2014 & 	56765.18459 &355.86  & G800L & 1880 &  -  \\ 

\enddata
\tablecomments{A log of  {\em HST} observations of GRB 130427A. Optical observations were obtained with ACS, while UV and IR imaging was taken with WFC3 in the UVIS and IR channels respectively. The magnitudes shown are for a point source
at the GRB location, and not for the overall host galaxy + afterglow combination. The magnitudes are corrected for
a foreground extinction of $A_{F336W} = 0.090$ mag, $A_{F606W} = 0.050$ mag and $A_{F160W}= 0.010$ mag. The late time
magnitudes refer to the flux measured at the GRB position, but are likely dominated by host galaxy light. Errors in the 
magnitudes are statistical only. }
\label{hst}
\end{deluxetable*}

\subsection{{\em Hubble Space Telescope} Observations}

We observed the location of GRB~130427A with {\em HST} on 20 May 2013, 23 days after the initial burst detection. A
second epoch was obtained in April 2014, almost a year after the initial burst. 
A log of these observations is shown in Table~\ref{hst}. 
For more detailed study of the host galaxy we also utilize a longer (2228 s) WFC3/F606W observation obtained
on  15 May 2014. The imaging data were reduced 
in the standard fashion; with on-the-fly processed data retrieved from the archive and subsequently re-drizzled using 
{\tt astrodrizzle}, for UVIS observations we separately corrected for pixel based charge transfer inefficiency (CTE) \citep{anderson12}. 
Photometry was performed in small apertures to minimize any contribution from underlying galaxy light, and maximise signal to noise, it was
subsequently corrected using standard aperture corrections\footnote{See \citealt{sirianni} for ACS aperture corrections 
and \url{http://www.stsci.edu/hst/wfc3/phot\_zp\_lbn} for WFC3}. We also use direct image subtraction to isolate the afterglow/SNe light
at early epochs, this effectively removes the host contribution. These magnitudes may still contain some transient light, but since the second
epoch magnitudes are a factor of 
$\sim$ 15-30 lower than observed at early times, this suggests that these epochs can be used for effective subtraction.The resulting photometry is shown in 
Table~\ref{hst}, while our {\em HST} images are shown in Figure~\ref{hst}. 

We also obtained grism spectroscopy centered at $\sim 8000$\AA\, with the G800L grism on ACS, with a
position angle chosen to minimize the contribution from the underlying host galaxy (see Figure~\ref{hst}). Again we utilized the on-the-fly calibrated images,
corrected for CTE and bias striping. We detected sources on a single F606W image, and extracted these via {\tt aXe}, subsequently drizzling
each of the four exposures to create master spectra, which was flux calibrated using published sensitivity curves. We
extracted the light from the GRB counterpart in a relatively small aperture ($\sim 2 \times FWHM$).
In principle a given pixel in the grism image
may be exposed to light of multiple different wavelengths from different spatial locations on the chip.
For the second epoch of observations we force an extraction of the same width
at the position of the transient (as determined by a map between the first and second epoch of direct imaging). We then subtract this
from the initial spectrum to obtain a host free spectrum. Since we have utilised a tight aperture around the SNe we subsequently scale
this subtracted spectrum to the host subtracted magnitudes of the afterglow/SNe.

\section{Isolating the supernova}

\subsection{Host contamination}

The afterglow is offset ($0.83 \pm 0.03$)\arcsec\, from the centroid of the host galaxy light in F606W, and so the latter is
of little concern in our small (0.1\arcsec) apertures. However, regions underlying GRBs are frequently 
amongst the most luminous parts of their hosts
\citep{fruchter06}, so some contamination may be expected. Our late time subtraction removes this from both our 
broad band photometry and grism spectroscopy. For the high resolution observations reported here this contamination is small
(at most 6\% in the UV and 3\% in the optical) for our photometry. However, the contribution is somewhat larger in the grism 
observations. These observations disperse light not only from the region directly underlying the GRB, but also from 
other locations in the host (which represent contributions at different wavelengths, overlapping the transient light).  In particular, the proximate,
bright star forming region contaminates the SNe considerably ($>$20\% at the red end of our spectrum beyond 9000\AA). This region is also likely to be the
dominant host contaminant in ground based spectroscopy (since the host galaxy is resolved), and has quite different colours
from the global host, implying potentially significant systematic errors when the broadband SDSS colours of the host
are used to attempt a subtraction (e.g. Xu et al. 2013, Melandri et al. 2014). Here we can directly remove this contribution via
the subtraction of the deep second epoch.

\begin{figure*}[ht]
    \centering
    \includegraphics[width=12cm,angle=0]{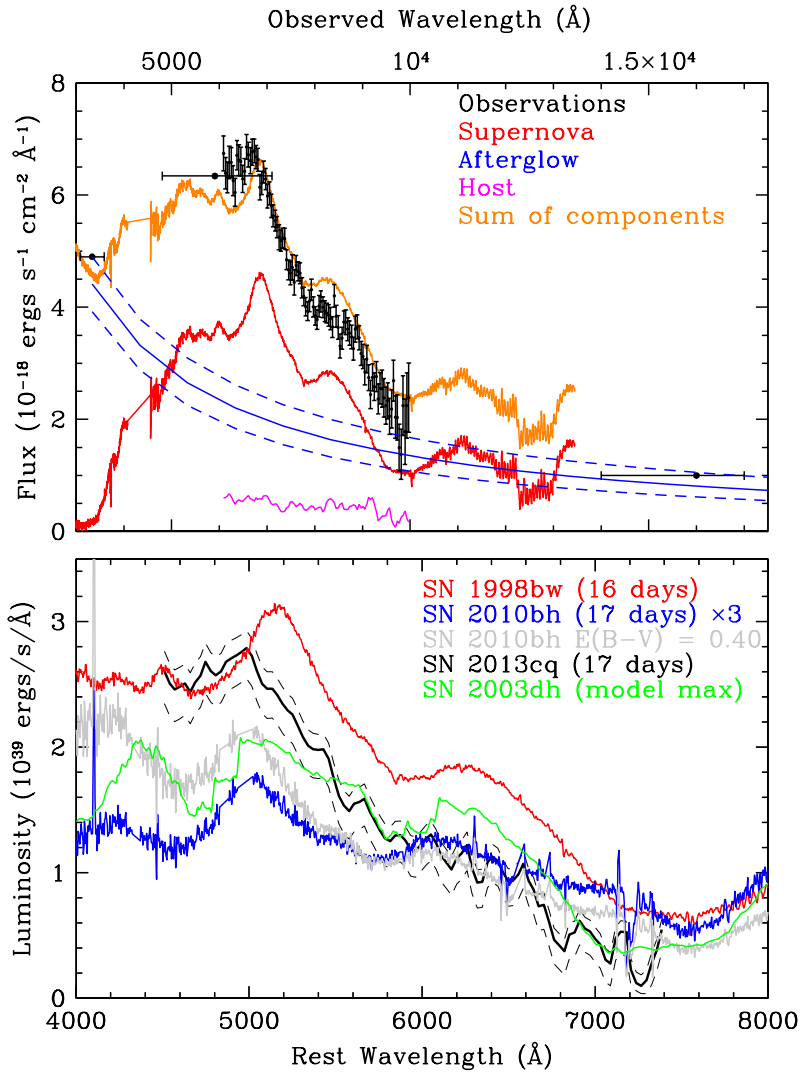}
\caption{The spectral energy distribution of GRB130427A/SN~2013cq as measured with {\em HST}. The top panel shows
the data (black) along with the different components that may
contribute as indicated. The host galaxy spectrum is based on our extraction of the host directly under the GRB position, 
and not its global properties. The lower panel shows the  smoothed SN spectrum
after subtraction of the afterglow light, and in luminosity space, directly compared against spectra of other GRB/SNe pairs. 
The supernovae have been scaled as shown in the legend, but in general the spectra show a good match with SN~1998bw
at a similar epoch.}
\label{SN}
\end{figure*}

\subsection{Afterglow}
SNe Ib/c are generally weak UV emitters due to the strong metal line blanketing shortward of $\sim 3000$\AA, so to first order 
our F336W observations should be free of supernova contribution. This is consistent with the UV colours of GRB060218/SN2006aj
 \citep[from][]{simon10} and of XRF100316D/SN2010bh \citep[from][]{cano11}, which would predict a factor $>10$ decrease in
 flux between F606W and F336W. 
There are few UV spectra of SN Ic, however if we graft the STIS UV observations of 
SN~2002ap onto the optical spectra of SN~1998bw following \citealt{levan2005b} then we can obtain a first order
approximation of the likely UV spectral shape
(see Figure~\ref{SN}).

These colours and spectra suggest it is reasonable to assume that the F336W light is dominated by the afterglow component. 
To confirm this we utilize the UV to IR lightcurve from
\citep{perley13}. This exhibits a spectral slope of $F_{\nu} \propto \nu^{\beta}$ with
$\beta \approx -0.92$ and predicts U(AB)=23.41 $\pm$ 0.10 at the time of the first epoch of {\em HST} F336W observations. 
The corresponding {\em HST} UV magnitude is F336W(AB)=23.28 $\pm$ 0.02. Corrected for foreground absorption this is consistent
with the afterglow contributing 90$\pm$10\% of the measured F336W flux, confirming our assumptions above. This afterglow
model predicts F160W(AB) = 21.84, somewhat fainter than the measured magnitude and suggesting that the supernova
makes up $\sim$20\% of the light in this band, again in keeping with expectations of the few IR spectra
of broad-lined SNe Ic obtained to-date \citep[e.g.][]{bufano2012}. 

We estimate the supernova contribution by using the 
model above, with initial error bars  accounting for the uncertainty in the 
F336W afterglow  light, discussed above,
 and the intrinsic value of $\beta$, which we adopt to be 0.92 $\pm 0.1$. For this
range of models we then subtract the afterglow spectrum from the measured grism data, and neglect any host
galaxy contamination. The extremum of this model is set by the assumption that both F336W and F160W are entirely
dominated by afterglow, and by subtracting the resulting power-law index. 

\section{Discussion}
\subsection{Supernova properties}
Our grism spectrum, both before and after subtraction of the afterglow and host
light, is shown in Figure~\ref{SN} (top and bottom). Broad features,
consistent with those seen in other high velocity SN Ic associated
with GRBs are clearly visible in the spectrum, even before subtraction
of the afterglow component.  The absence of broad emission at
H$\alpha$ or He absorption rules out type II or Ib events respectively. 
In the lower panel of Figure~\ref{SN}
we plot rest-frame wavelength versus luminosity comparisons of the
afterglow subtracted and de-reddened\footnote{We de-redden the
spectra with a \citealt{fitzpatrick99} law, for a Galactic E(B-V) = 0.02 and intrinsic E(B-V) =0.05, consistent
with the Na {\sc i} D Doublet \citep{xu13} and afterglow modelling \citep{perley13,maselli14}. The comparison
SNe have been de-reddened utilising E(B-V) = 0.07 for SN~1998bw, and a range of E(B-V) for SN~2010bh. The
spectrum of SN~2003dh is from a spectral model free from extinction.} spectra of
SN~2013cq with various GRB/SNe pairs.

The 
closest match for the overall spectral shape and luminosity
is that of SN~1998bw \citep{galama98,iwamoto98,patat01}. 
 The
similarity in appearance of these SNe is primarily due to the overall spectral shape, with
a drop in luminosity of a factor of $\sim$ 3 over the 5000-7000\AA~ range,
substantially more than seen in other GRB/SNe pairs. 

The broad colours of these SNe are similar, and if the kink at $\sim 6000$\AA~ is
interpreted as the SiII (6355\AA) blend then it would be indicative
of a photospheric velocity similar to SN~1998bw at the same epoch 
($v_{ph} \sim 15000$ km s$^{-1}$),  although we note
that this feature is apparently stronger in SN~1998bw, where there is marked
upturn in flux redward of it. This may suggest
a somewhat higher velocity for SN~2013cq . Taken at face value
this would suggest that SN~2013cq is broadly similar in peak luminosity,
$^{56}$Ni production and kinetic energy to SN~1998bw.  

While the similarity to SN~1998bw is very marked
over the rest frame spectral range from 5100-7000\AA~
it is much poorer around the central peak of an SN Ic at $\sim 5000$\AA. 
In SN~2013cq this feature appears to be broader, and blueshifted relative
to SN~1998bw, as observed by \citealt{xu13}. Our comparison suggests
that this peak may fit better with the slightly higher velocity SN 2003dh \citep{hjorth03,mazzali03},
the only one of the comparison SNe to arise from a luminous
cosmological GRB.  In this case, the SN~2003dh model shown in Figure 2 remains
a factor of 50\% less luminous than SN~2013cq, and is systematically redder (i.e. a smaller
decrement between the peak and $\sim 6000\AA$). However, the positions of the broadened
lines do then appear generally similar.

Alternatively, this could be suggestive
of a much higher velocity supernova such as SN~2010bh \citep{bufano2012,cano11} as favoured
by \citealt{xu13}, who infer $v_{ph}=32,000$ km s$^{-1}$ from the  FeII (5169\AA) at an epoch of 12.5 rest-frame days.
This does provide a good match to the location and width of the feature;
however, the overall spectral shape and luminosity of SN~2010bh are also very different from 
SN~2013cq. At 17 days SN~2013cq appears to be much bluer, and a factor $>$3 times more
luminous than SN~2010bh, suggesting it is not as good
an analog as SN~1998bw. 
To obtain a match in both luminosity and general spectral shape would require that
the reddening in the direction of SN~2010bh had been significantly underestimated. 
Our plotted spectrum assumes $E(B-V)_{gal} = 0.12$ and $E(B-V)_{int}$ = 0.14 
\citep{bufano2012}. To obtain a match would require de-reddening the spectra with
an {\em additional} $E(B-V)=0.4-0.5$, well beyond the values allowed from the Na {\sc i} D doublet
observed in moderate resolution X-shooter spectroscopy \citep{bufano2012} 
and even the largest extinctions allowed by the afterglow spectrum \citep{olivares}.  Hence,
we disfavour the suggestion that the underlying SNe is similar to SN~2010bh, due to the
significant disparity in the luminosity. Instead we favour a SNe similar in $^{56}$Ni yield
and kinetic energy to SN~1998bw and SN~2003dh, in which possible velocity structure within the ejecta
gives rise to a range of features within the spectra that do not provide a unique match
to any previous GRB/SNe pair. Indeed, the recent study of \citealt{melandri14} also identifies a source
which is much brighter than SN~2010bh, but spectrally intermediate between SN~1998bw and 
SN~2010bh.

Despite these differences, it is clear that the features observed in the spectrum 
of SN~2013cq are broadly compatible to the range observed in SN/GRB
pairs, although a full model description of the spectrum is beyond the scope of this paper. 
To obtain approximate bounds on the luminosity of
the supernova we integrate our afterglow and host subtracted spectra (allowing
for the errors in the afterglow parameters) through 
through a rest-frame V-band
filter to obtain a luminosity relative to SN~1998bw. This suggests that SN~2013cq 
has a luminosity factor $k=L_V (\mathrm{2013cq}) /L_V(\mathrm{1998bw})$
of $0.77 \pm 0.10$. 
Given the suggestions that
SN~2013cq is more rapidly evolving than SN~1998bw \citep{xu13},
these observations
(at the V-band peak of SN~1998bw) may
underestimate by true peak luminosity, which may be even closer to
that of SN~1998bw. 

\begin{figure}[ht]
    \centering
    \includegraphics[width=8cm,angle=0]{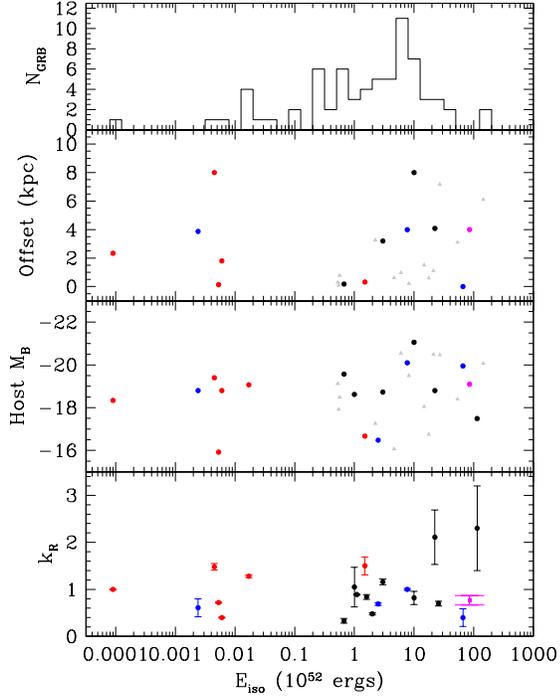}
\caption{Top: A histogram of $E_{iso}$ for {\em Swift} GRBs \citep[from][]{kocevski08}, amended with
the $E_{iso}$ values for GRB/SN pairs when not from {\em Swift}. 
Bottom: The R-band luminosities of candidate GRB/SN (scaled to SN~1998bw, which has $k_R = 1$), against the 
isotropic luminosities of the GRBs. Points in red are those with strong spectroscopic evidence for associated
supernovae (category A in \citealt{hjorth11}), blue points indicate cases where spectral features were seen at lower signal to noise
(category B in \citealt{hjorth11}), and black
points are those with weaker evidence for associated SNe. The two errors bars marked in the case of SN~2013cq represent the
error associated with a simple afterglow subtraction and the extrema of possibilities (see text for details). 
The middle two panels show the offset from the GRB host centre and the GRB host B-band absolute magnitude
as a function of isotropic energy release. The colour coding is the same as for the lower plot, while grey triangles
indicate values from the literature for GRBs without claimed SNe associations (data from \citep{bloom02,frail01}). These
data show that the properties of GRB SNe and host galaxies are largely unaffected by the energies of the burst, and hence
that progenitors in similar environments and with similar initial masses can likely create the entire GRB luminosity function. }
\label{eiso}
\end{figure}

GRB~130427A is unusual as a luminous GRB at a redshift
at which the supernovae are open to detailed study. Indeed, 
it the the highest luminosity burst for which there is spectroscopic
evidence for a supernova. 
In this regard the similarity of the supernova
to that seen in a burst (GRB 980425) that was six orders of magnitude
less energetic is striking. As we show in Figure~\ref{eiso} there
is no correlation between the GRB energetics and the
inferred peak magnitudes of their SNe. 
These similarities in SNe peak luminosity, and in their spectra suggest similarities in
the ejected  $^{56}$Ni masses and kinetic energies. Recent modelling
\citep{mazzali13} has used these diagnostics to suggest 
that most broad-lined SN Ic associated
with low luminosity GRBs in turn arise from a relatively small range of 
ZAMS progenitor masses, perhaps between $30-50$ M$_{\odot}$. The
detection of a similar SN in a highly luminous GRB extends this
across a broader range of energy.

\subsection{Host galaxy}

Our {\em HST} observations clearly resolve the afterglow from the host galaxy, showing the GRB to lie at a spatial offset of 0.83\arcsec,
(4.0 kpc at $z=0.34$) from the centre of its host. This is a relatively large offset for a GRB from its host galaxy ($>75$\% of those in
\citealt{bloom02}) although by no means exceptional (see Figure 3). 
In the F606W observations the galaxy exhibits a bar-like structure with a face-on
disk visible beyond this, a weak spiral structure is also apparent, making the host one of few to be classified as a spiral  \citep{svensson2010}. 
The F336W imaging shows weak star formation close to the centre
of the galaxy, but the most striking feature is a strong star forming region at a similar radial offset to the GRB, but offset from the
GRB position by 0.3\arcsec (1.5 kpc).  This region the strongest region of star formation in the host galaxy. While the
magnitude of star formation underlying the GRB position remains uncertain because of contamination with the late time afterglow, the
offset region is at least a factor of $\sim 2$ brighter. It also shows readily detectable OIII emission in the late time grism spectrum. 
The weak spiral arm in the direction of the GRB also appears distorted, and so it may be that star formation
has been triggered via a tidal interaction. In the larger field around GRB~130427A we note several galaxies of similar magnitude
(see Figure~\ref{hst}). In our grism spectroscopy these galaxies do not exhibit strong spectral lines, but they may well imply
that the host of GRB~130427A lies within an association. 

The magnitude of the galaxy in a large aperture  at late times after the afterglow/SNe has significantly faded is
F336W(AB)=22.84 $\pm$ 0.07
 at a rest-frame wavelength of approximately 2500\AA. This corresponds to a UV
star formation rate of $\sim 1.1 \pm 0.1 $ M$_{\odot}$ yr$^{-1}$ (error statistical only), broadly in agreement with that inferred from
the SDSS observed SED of the galaxy of $2^{+5}_{-1}$ M$_{\odot}$ yr$^{-1}$ \citep{xu13}, and suggesting
a relatively (but not extremely) low specific star formation rate by the standards of GRB hosts \citep{svensson2010}. The compact star forming region close to the GRB
contains $\sim$10\% of this star formation, making a highly luminous star forming region comparable to that seen in the
host of GRB 980425/SN~1998bw \citep{hammer06,christensen08} or in a handful of other local galaxies \citep{beck96}.  At late times UV
emission is visible close to the GRB position, indicating that lower level star formation is likely arising proximate to the GRB. This may still contain some
afterglow light, but this region is at least a factor $>1.5$ fainter than the brightest star forming region in the host.

\begin{figure*}[ht]
    \centering
    \includegraphics[width=11cm,angle=270]{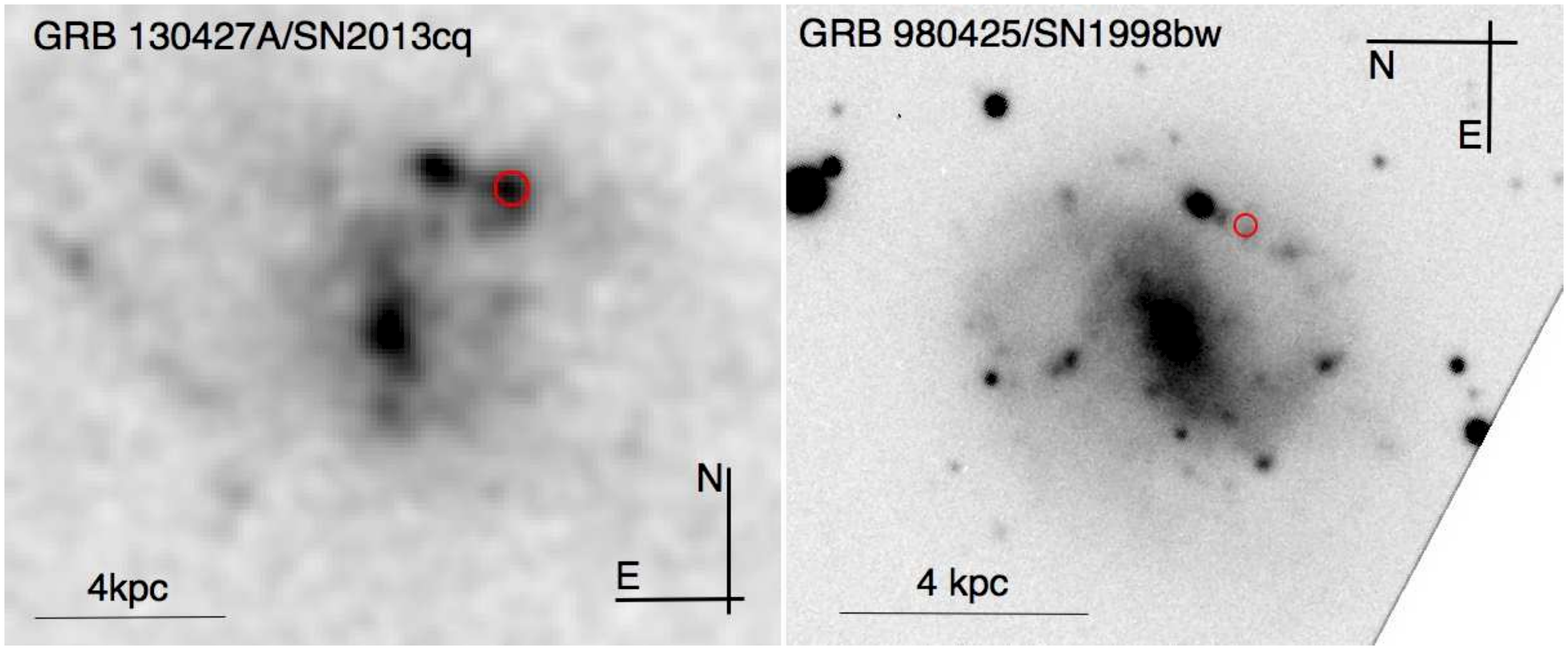}
\caption{The host galaxies of GRB 130427A (left) and GRB 980425 (right). The left hand image is taken from our {\em HST} observations in F606W (May 2014), while the right
hand image is an archival image of the host obtained with VLT/FORS2 in 2004. The position of the GRB/SNe is marked with a red circle in each case (note that the GRB 130427A image may contain afterglow contribution at the GRB/SN position). The resemblance between the two host galaxies is striking, especially given the rarity of spiral hosts amongst GRB host galaxies. Interestingly unlike most bursts these do not lie on the brightest regions of the hosts, but in spiral arms where the metallicity may be lower (see Fruchter et al. 2006). In both GRB 130427A and GRB 980425 the GRB/SNe occurred in a relatively faint region within the spiral arm, and not in the strongest star forming complex.   } 
\label{98bwcomp}
\end{figure*}

Some previously extremely energetic bursts (e.g. GRB 080319B, \citealt{tanvir10}) have shown extremely faint and small host galaxies, while the
hosts of GRBs 980425, 030329 and 060218 are also sub-luminous and LMC-like. However, as show in Figure~3 the overall population of
GRB-SN host luminosities is 
shows no discernible correlation with
the energy of the GRB (see also \citealt{levesque2010}), and the host of GRB~130427A is in keeping with these expectations.  It is intriguing to note
that the GRB host galaxy with the closest properties to that of GRB 130427A/SN~2013cq is in fact the host of SN~1998bw. It is also a rare example
of a spiral galaxy, in which the GRB occurs at a moderately large offset from the nucleus, and from the strongest region of star formation within the host galaxy. 
This is shown graphically in Figure~\ref{98bwcomp}. Given the similarities in supernova and environment GRB~130427A would seem like a close analog
of GRB 980425A if it were not for the factor of $10^6$ difference in their $\gamma$-ray energy releases.

\section{Conclusions}
We have presented {\em HST}  imaging and spectroscopic observations of the extremely bright GRB 130427A, which
show it was associated with a luminous broad line SN Ic (SN 2013cq). The red spectra offer good agreement with those of
SN~1998bw, while the bluer spectra appear well matched in position if not shape with SN~2010bh.  The host galaxy appears
to be a disk galaxy of moderate luminosity and star formation rate, whose overall characteristics are consistent with those
of the GRB host population at large. The similar properties
of the SNe and hosts over six orders of magnitude in GRB isotropic equivalent energy
would appear to suggest
that the energy of the GRB is not a strong function of environment or the mass of
the progenitor star, and that stars of similar mass and composition
are responsible for the entire luminosity function of GRBs. More
complex effects within the star (e.g. rotation) or geometric effects
are therefore needed to explain much of the diversity in the GRB luminosity
function.

\section*{Acknowledgements}

We thank Matt Mountain and the STScI staff for rapidly scheduling our observations. AJL thanks the Leverhulme Trust. AJL, NRT and KW are supported by STFC. The Dark Cosmology Centre is funded by the DNRF. Based on observations made with the NASA/ESA Hubble Space Telescope, obtained at the Space Telescope Science Institute, which is operated by the Association of Universities for Research in Astronomy, Inc., under NASA contract NAS 5-26555. These observations are associated with program \# 13230, 13110 \& 13117.


\end{document}